\documentclass[aps,prl,twocolumn,groupedaddress,showkeys,showpacs]{revtex4}

\usepackage{amssymb,amsbsy,amsmath}
\usepackage{multirow}
\usepackage{graphicx}

\newcommand{\xref}[1]{\protect\ref{#1}}
\newcommand{\figref}[1]{Fig.~\protect\ref{#1}}
\newcommand{\fmref}[1]{(\protect\ref{#1})}

\newcommand{\bs}{\boldsymbol}
\begin{document}

\title{Heat capacity uncovers physics of a frustrated spin
  tube}

\author{Nedko B. Ivanov$^{1,2}$}
\email{<Nedko.Ivanov@Physik.Uni-Magdeburg.DE>}
\author{J\"urgen Schnack$^{2}$}
\email{jschnack@uni-bielefeld.de}
\author{Roman Schnalle$^{2}$}
\author{Johannes Richter$^{3}$}
\author{Paul K\"ogerler$^{4}$}
\author{Graham N. Newton$^{5}$}
\author{Leroy Cronin$^{5}$}
\author{Yugo Oshima$^{6}$}
\author{Hiroyuki Nojiri$^{6}$}
\email{nojiri@imr.tohoku.ac.jp}

\affiliation{$^1$Institute of Solid State Physics, Bulgarian Academy
  of Sciences, Tzarigradsko chaussee 72, 1784 Sofia, Bulgaria}
\affiliation{$^2$Fakult\"at f\"ur Physik, Universit\"at
  Bielefeld, Postfach 100131, D-33501 Bielefeld, Germany}
\affiliation{$^3$Institut f\"ur Theoretische Physik, Universit\"at Magdeburg,
P.O. Box 4120, D-39016 Magdeburg, Germany}
\affiliation{$^{4}$Institut f{\"u}r Anorganische Chemie, RWTH
  Aachen, Landoltweg 1, D-52074 Aachen, Germany}
\affiliation{$^{5}$Dept. of Chemistry, The University of Glasgow, Glasgow, G12 8QQ, UK}
\affiliation{$^{6}$Institute for Materials Research, Tohoku
  University, Katahira 2-1-1, Sendai 980-8577, Japan}

\date{\today}

\begin{abstract}
We report on refined experimental results concerning the
low-temperature specific heat of the frustrated spin tube
material [(CuCl$_2$tachH)$_3$Cl]Cl$_2$. This substance turns out
to be an unusually perfect spin tube system which allows to
study the physics of quasi-one dimensional antiferromagnetic
structures in rather general terms. An analysis of the specific
heat data demonstrates that at low enough temperatures the
system exhibits a Tomonaga-Luttinger liquid behavior
corresponding to an effective spin-3/2 antiferromagnetic
Heisenberg chain with short-range exchange interactions. On the
other hand, at somewhat elevated temperatures the composite spin
structure of the chain is revealed through a Schottky-type peak in
the specific heat located around $2$~K.  We argue that the
dominating contribution to the peak originates from gapped
magnon-type excitations related to the internal degrees of
freedom of the rung spins.
\end{abstract}

\maketitle
%%%%%%%%%%%%%%%%%%%%%%%%%%%%%%%%%%%%%%%%%%%%%%%%%%%%%%%%%%%%%%%%%%%%%%%%

\begin{figure}[ht!]
\centering
\includegraphics*[clip,width=65mm]{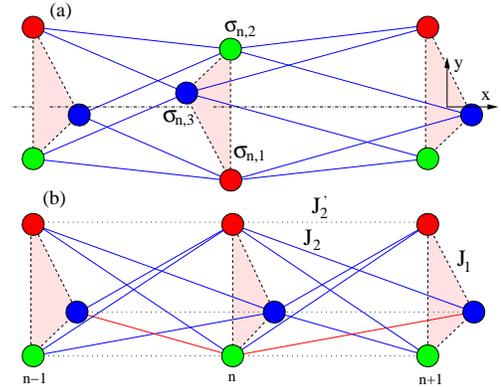}
\caption{(a) Sketch of the twisted spin-tube system. (b) An equivalent
  spin model obtained through inversion of every second
  triangle. The $n$th elementary cell contains the spin-1/2
  operators $\bs{\sigma}_{n,1},\bs{\sigma}_{n,2}$, and
  $\bs{\sigma}_{n,3}$.} 
\label{F-A}
\end{figure}

\emph{Introduction}--Spin tubes constitute a special class of
quasi-one dimensional spin ladder systems characterized by
periodic boundary conditions in the rung direction
\cite{MHF:JSSC99,GRM:PRL03,LNM:PRB04,Sat:JPCS05,Sat:PRB05,MSS:PE05,SMO:PE05,
SVC:PRL05,OYS:PTPS05,FLP:PRB06,SaS:PRB07,NiA:PRB08,ZGS:PRB08}.
The magnetic compound [(CuCl$_2$tachH)$_3$Cl]Cl$_2$ is a
geometrically frustrated triangular spin tube, the frustration
being related both to the triangular arrangement of its rungs
and to the twisted geometry of the legs, compare
\figref{F-A}(a).  The relatively simple exchange pathway
structure, described only by two dominant Heisenberg exchange
couplings \cite{SKK:CC04,SNK:PRB04}, as well as the extremely
weak exchange interactions between neighboring tubes renders
[(CuCl$_2$tachH)$_3$Cl]Cl$_2$ an excellent real material to
study general properties of the spin-tube systems
\cite{Sat:PRB05,SMO:PE05,FLP:PRB06,Sch:CRC07,NiA:PRB08}. An
appropriate spin Hamiltonian describing the magnetic properties
of this material reads as
%--------------------------------------------------------
\begin{eqnarray}
\label{E-2-1}
{\cal H}&=&\sum_{n=1}^L\sum_{\alpha=1}^3\left[
J_1\,\bs{\sigma}_{n,\alpha}\cdot\bs{\sigma}_{n,\alpha+1}\right.
\\
&+&\left.
J_2\,\bs{\sigma}_{n,\alpha}\cdot\left( 
\bs{\sigma}_{n+1,\alpha+1}+\bs{\sigma}_{n+1,\alpha-1}
\right) \right]
\nonumber
\ ,
\end{eqnarray}
%--------------------------------------------------------
where $\bs{\sigma}_{n,\alpha}$ ($\alpha=1,2,3$) are spin-1/2
operators defined on the vertices of the elementary triangle
denoted by index $n$ ($n=1,\ldots,L$).  As depicted in
\figref{F-A}~(b), the twisted spin tube may also be thought of
as a three-leg ladder with periodic boundary conditions in the
rung direction, where the parameters $J_1$ and $J_2$ are
strengths of the rung and crossing (diagonal) exchange bonds,
respectively. [(CuCl$_2$tachH)$_3$Cl]Cl$_2$ is characterized by
the parameters $J_1/k_B=1.8$ K and $J_2/k_B=3.9$ K
\cite{SNK:PRB04}, whereas the leg exchange constant $J_2^{'}>0$,
introduced for the sake of clarity in \figref{F-A}~(b), seems to
vanish in this material.  Figure~\ref{F-A}~(b) clearly reveals
the translation symmetry with one triangle per unit cell.

%===================    figure   =================================
%===================    figure   =================================

Two extreme scenarios for the ground state of Eq.~(\ref{E-2-1})
with antiferromagnetic couplings ($J_1,J_2>0$) were outlined in
Ref.~\cite{FLP:PRB06}. In the case of dominating $J_1$
couplings, the system effectively maps onto an effective
spin-chirality model, where the additional chirality degrees of
freedom appear as a result of the ground-state degeneracy of
each individual triangle. On the other hand, for dominating
$J_2$ couplings the system maps onto an effective spin-3/2
antiferromagnetic Heisenberg chain (AHC) containing additional
biquadratic exchange couplings. In this Letter, we demonstrate by
means of refined specific heat measurements, that the
low-temperature properties of the spin tube material
[(CuCl$_2$tachH)$_3$Cl]Cl$_2$ reproduce the behavior of a
spin-3/2 AHC characterized by the effective short-range exchange
coupling constant $J_{\text{eff}}=2J_2/3$.  Since the exchange
interactions between different tubes are extremely small, the
discussed compound also provides a rare example of spin-3/2 AHC
\cite{Xia:PRB98,IEK:PRB99,PhysRevLett.76.4955,KlJ:PRL00,PhysRevB.61.9558}.
The experimental observables strongly suggest that if the system
ever orders this should be much below 0.1~K.  At elevated
temperatures the measured specific heat exhibits a big
Schottky-type peak located around $T\approx 2$ K.  A detailed
analysis -- combining the semiclassical spin-wave approach with a
number of numerical techniques such as the Quantum Monte Carlo
(QMC) method, the Lanczos exact numerical diagonalization (ED),
and the complete exact diagonalization
\cite{SHS:JCP07,ScS:PRB09} -- implies that the main contribution
to the specific-heat peak stems from the lowest-lying gapped
magnon excitations resulting from the internal degrees of
freedom of the composite rung spins.

For the following discussions it is instructive to rewrite
Eq.~\fmref{E-2-1} in the form 
%--------------------------------------------------------
\begin{eqnarray}
\label{E-2-2}
{\cal H}
=\sum_{n=1}^L\left[\frac{J_1}{2}
\left(\bs{s}_n^2-\frac{9}{4}\right)+J_2\,\bs{s}_n\cdot\bs{s}_{n+1}\right]
+V\ .
\end{eqnarray}
%--------------------------------------------------------
Here
$\bs{s}_n=\bs{\sigma}_{n,1}+\bs{\sigma}_{n,2}+\bs{\sigma}_{n,3}$
is the rung spin operator related to the $n$th triangle.  The
number of Cu sites is $N=3L$, where $L$ is the number of rungs.
The interaction term $V$ reads as
%--------------------------------------------------------
\begin{eqnarray}
\label{E-2-3}
V=(J_2^{'}-J_2)
\sum_{n=1}^L\sum_{\alpha=1}^3\bs{\sigma}_{n,\alpha}\cdot
\bs{\sigma}_{n+1,\alpha}
\ ,
\end{eqnarray}
%--------------------------------------------------------
where we have included an additional leg exchange interaction
$J_2^{'}$, see \figref{F-A}~(b).  The last two equations
explicitly show that the condition $J_2=J_2^{'}$ defines a
special symmetric line in the parameter space on which the rung
spin operators $\bs{s}_n$ are conserved quantities. On this
line, the Hilbert space is decomposed into the invariant
subspaces (sectors) $[s_1,s_2,\ldots,s_L]$, where the local spin
quantum numbers $s_n=1/2,3/2$ are defined, as usual, by the
relations $\bs{s}_n^2=s_n(s_n+1)$ ($n=1,\ldots ,L$). In
particular, if the Hamiltonian ground state lies in the sector
$[3/2,3/2,\ldots, 3/2]$, then Eq.~\fmref{E-2-2} for
$J_2=J_2^{'}$ will describe a spin-3/2 AHC characterized by the
effective exchange constant $J_{\text{eff}}=J_2$.  Up to first
order in the (formally) small parameter $(J_2^{'}-J_2)/J_1$, the
effect of the interaction term $V$ in Eq.~(\ref{E-2-2}) is
reduced to a simple renormalization of the exchange constant in
the effective spin-3/2 AHC: $J_{\text{eff}}\longrightarrow
J_{\text{eff}}=J_2+(J_2^{'}-J_2)/3$. At the special point
$J_2^{'}=0$, the result $J_{\text{eff}}=2J_2/3$ coincides with
the first-order result of Ref.~\cite{FLP:PRB06}, which is
obtained by another perturbation scheme starting from the limit
$|J_1|\gg J_2$ ($J_1<0$).

\emph{Low-lying spin excitations}--As argued below, the
contributions to the low-temperature specific heat of the spin
tube material [(CuCl$_2$tachH)$_3$Cl]Cl$_2$ are dominated by
three types of low-lying spin excitations.  Apart from the
standard gapless excitations (characteristic of any half-integer
AHC with short-range exchange couplings), important
contributions to the specific heat appear from two additional
branches of low-lying gapped magnon-type modes, which are
related to the chirality degrees of freedom of the local
triangles.
 
A qualitative picture of the low-lying spin excitations can be
obtained already in the framework of the semiclassical spin-wave
approach starting from the classical N\'{e}el configuration
$|S_t,-S_t,\ldots \rangle$, where $S_t$ is the maximal value of
the $z$-component of the rung spin ($S_t=3/2$ in the present
case).  Since the elementary cell contains three spin-$S$
variables, there appear three different branches of spin-wave
modes
%--------------------------------------------------------
\begin{eqnarray}
\label{E-4-2}
&&E_{m}(k_x)=4SJ_2\times 
\\ \nonumber
&&\sqrt{\left(1\! -\! \alpha \sin^2 \frac{k_y}{2}\right)^2 \!
-\! \frac{1}{4}\left[\cos k_x\! +\! \cos (k_x\! +\!
k_y)\right]^2}
\ ,
\end{eqnarray}
%--------------------------------------------------------
where $\alpha = J_1/J_2$, $k_y=2\pi m /3$ ($m=0,1,2$), and $S$
is the spin of a single site ($S=1/2$ in the present case).  As
may be expected, the energy of the $m=0$ branch does not depend
on the parameter $J_1$, since it is related to the dynamics of
the cell spins ${\bs s}_n$ as a whole:
%--------------------------------------------------------
\begin{eqnarray}
\label{E-4-3}
E_0(k_x)=v_s|\sin k_x| ,
\hspace{1cm} v_s=(6S)\frac{2J_2}{3}
\ .
\end{eqnarray}
%--------------------------------------------------------
The above expressions reproduce the well-known semi-classical
results for the dispersion relation and the related spin-wave
velocity $v_s$ of a spin-($3\cdot S$) AHC with the effective
exchange constant $J_{\text{eff}}=2J_2/3$.

%===================    figure   =================================
\begin{figure}[ht!]
\centering
\includegraphics*[clip,width=65mm]{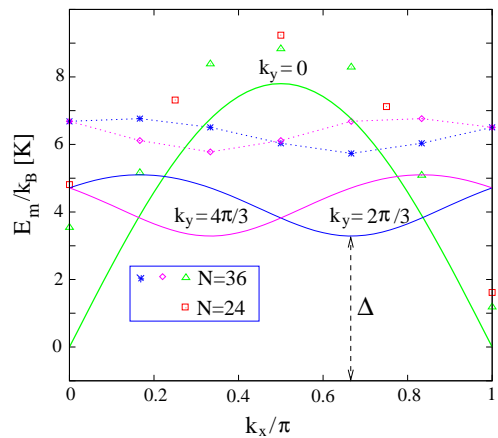}
\caption{Spin-wave excitation modes (solid curves,
  Eqs.~\fmref{E-4-3} and \fmref{E-4-4}) compared with the lowest
  triplet excitations in two periodic clusters (symbols).}
\label{F-B}
\end{figure}
%===================    figure   =================================

The dispersion relations  of the $k_y=2\pi/3$ and $k_y=4\pi/3$ 
excitations can be expressed in the following form
%--------------------------------------------------------
\begin{eqnarray}
\label{E-4-4}
E_{1,2}(k_x)=\sqrt{\Delta^2+4S^2J_2^2\sin^2 \left(
k_x\mp\frac{2\pi}{3}\right)}
\ ,
\end{eqnarray}
%--------------------------------------------------------
where the excitation gap $\Delta =
4SJ_2\sqrt{(1-3\alpha/4)^2-1/4}$ corresponds to the lowest-lying modes
at wave vectors $k_x=\pi/3$ and $2\pi/3$.

The solid curves in Fig.~\xref{F-B} show the dispersion
relations of the discussed spin-wave excitations. On the other
hand, the symbols depict the positions of the lowest-lying triplet
states, as obtained by the ED method for periodic clusters
containing $L=8$ and $12$ unit cells \footnote{Apart from the
  discussed two branches of gapped 
triplet modes, there are two additional branches of gapped
singlet modes also related to the chirality degrees of freedom
of the local triangles. However, in the experimentally
interesting region of the phase diagram ($J_1/J_2=0.46$) these
singlet modes lie above the triplet excitations presented in
\figref{F-B}.}.
Apart from the
finite-size effects related to the ED results, it is clearly
seen that both methods qualitatively yield similar results. As a
matter of fact, the discussed spin-wave branches may also be
considered as one-dimensional analogs of the three spin-wave
branches in the triangular lattice antiferromagnet.  In this
respect, the lowest-lying excitations at $k_x=0, 2\pi/3$ and
$4\pi/3$ in the spin tube are one-dimensional analogs of the
three Goldstone modes in the triangular lattice antiferromagnet.

\begin{figure}[ht!]
\centering
\includegraphics*[clip,width=38mm]{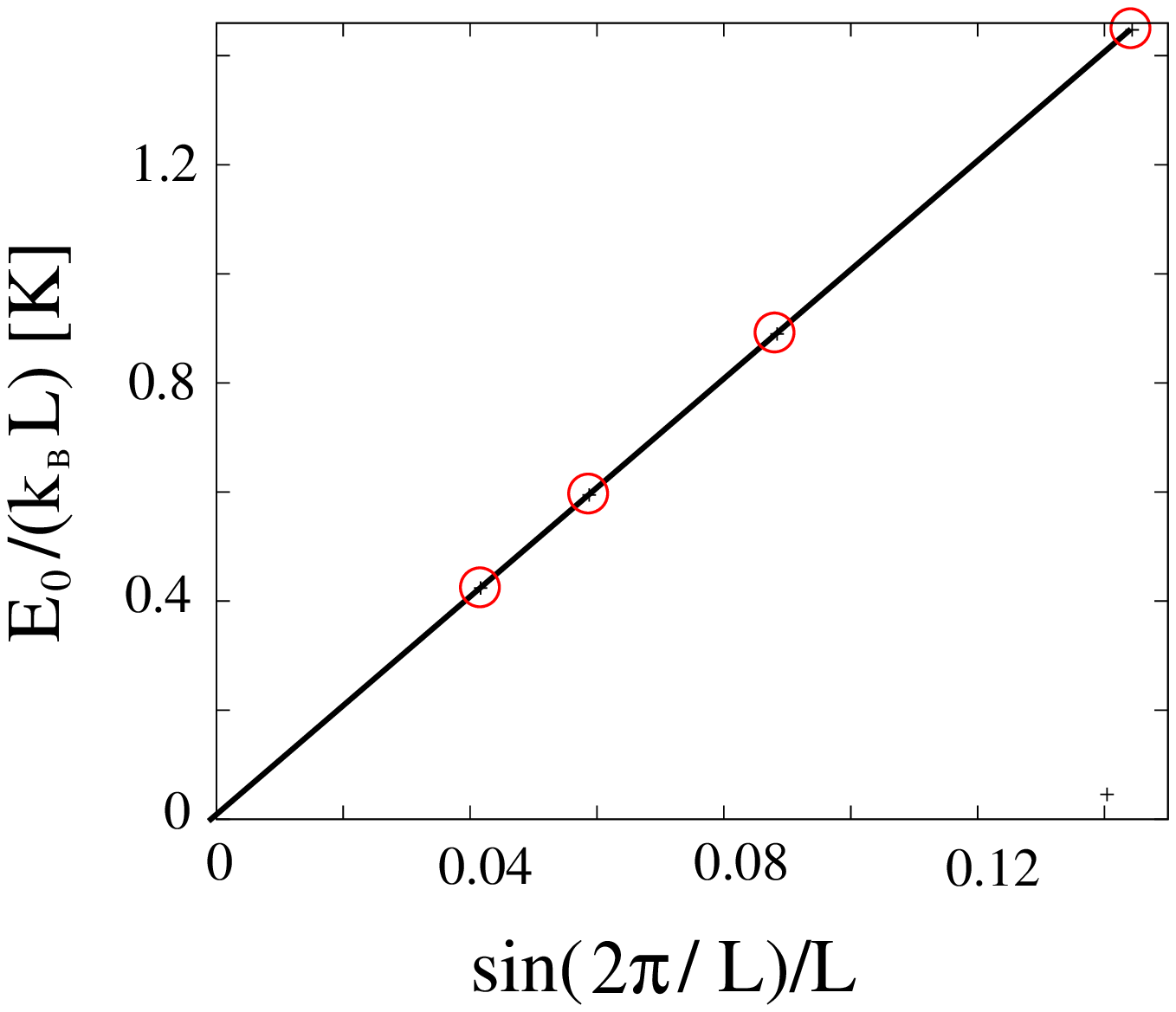}
\includegraphics*[clip,width=38mm]{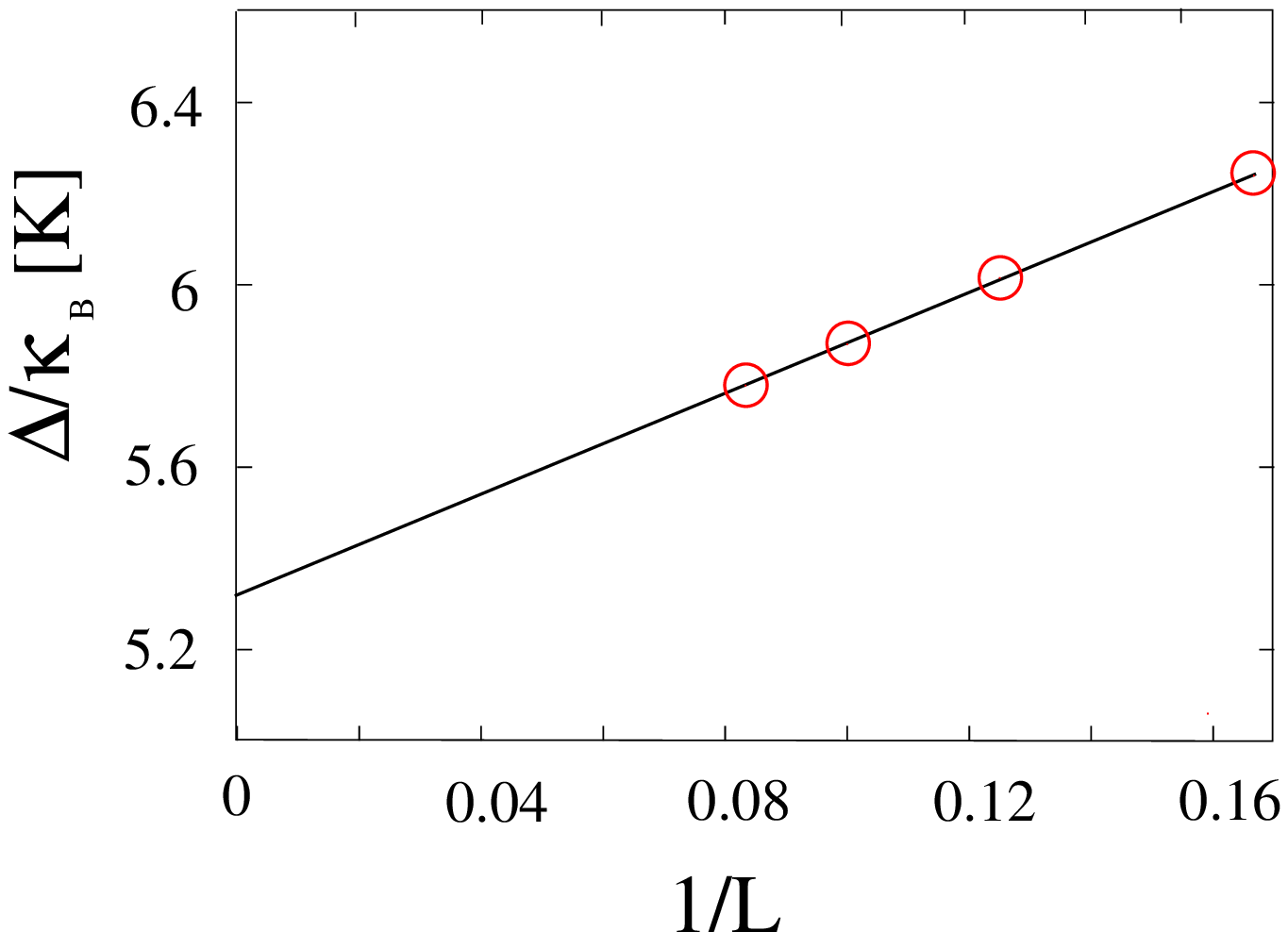}
\caption{Extrapolation of the ED results giving the parameters
of the excitation spectrum $v_s$ (l.h.s.) and $\Delta$ (r.h.s.).}
\label{F-E}
\end{figure}

More accurate estimates for the parameters of the excitation
spectrum $v_s$ and $\Delta$  can be
obtained from an extrapolation of  the ED results  for 
$L=6,8,10$, and $12$ unit cells.  Using  the  
approach  of Ref.~\cite{PhysRevLett.76.4955}, 
one finds the following estimates from the extrapolations of  
$E_0(2\pi /L)$ $vs.$ $\sin (2\pi/L)/L$ and
$E_1(2\pi /3)$ $vs.$ $1/L$ (see \figref{F-E}):
%--------------------------------------------------------
\begin{eqnarray}
\label{E-4-5}
v_s/k_B=10.06\,  \rm{K}= 3.87\, \frac{2J_2}{3k_B}
\ , \,\,\Delta/k_B=5.32\,  \rm{K} 
\ .
\end{eqnarray}
%--------------------------------------------------------
Interestingly, the  extrapolation result $3v_s/(2 J_2)=3.87$  exactly 
reproduces the 
density-matrix renormalization group estimate for the spin-wave 
velocity of the spin-3/2 AHC characterized
by the exchange constant $2J_2/3$ \cite{PhysRevLett.76.4955}. 
As already discussed, the same effective 
exchange  constant ($J_{\text{eff}}=2J_2/3$) arises both in  the 
first-order result for the effective spin model
and in the semiclassical spin-wave approach. Below we demonstrate 
numerically that the specific heat of a spin-3/2 AHC 
with the  exchange constant $2J_2/3$  excellently reproduces 
the experimental results in the low-temperature region $T\leq 0.5$~K.

%===================    figure   =================================
%===================    figure   =================================

\begin{figure}[ht!]
\centering
\includegraphics*[clip,width=65mm]{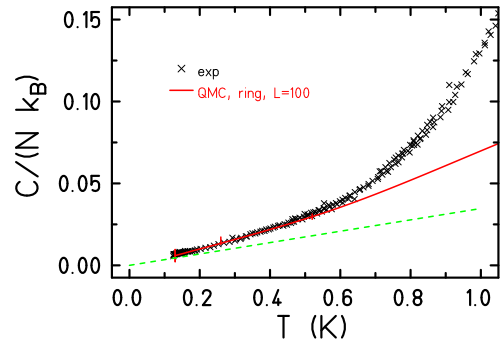}
\caption{Specific heat (per Cu spin) of
  [(CuCl$_2$tachH)$_3$Cl]Cl$_2$. The symbols denote the
  experimental values. The solid curve
  is the QMC result for a spin-3/2 chain of length $L=100$. The
  dashed line provides the linear specific heat corresponding
to the universal Tomonaga-Luttinger liquid form presented by
Eq.~(\ref{E-4-1}), by using the extrapolation result
  $v_s=10.06$~K.} 
\label{F-C}
\end{figure}

\begin{figure}[ht!]
\centering
\includegraphics*[clip,width=65mm]{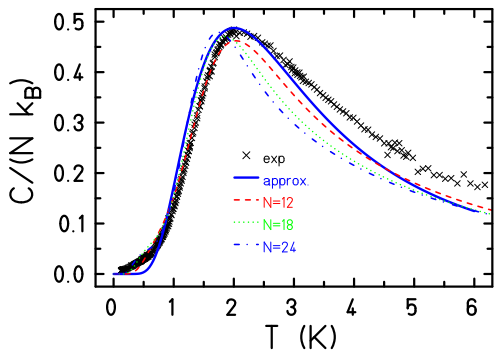}
\caption{Specific heat (per Cu spin) of
  [(CuCl$_2$tachH)$_3$Cl]Cl$_2$. The symbols denote the
  experimental values, whereas the solid curve depicts the
  two-level approximation. The broken curves denote the specific
  heat for three complete diagonalizations for finite sizes.}
\label{F-D}
\end{figure}

%%%%%%%%%%%%%%%%%%%%%%%%%%%%%%%%%%%%%%%%%%%%%%%%%%%%%%%%%%%%%%%%%%%%%%%%
\emph{Low-temperature specific heat}--The specific heat,
Figs.~\ref{F-C} and \ref{F-D}, was measured at the Institute for
Materials Research (IMR) of Tohoku University using
polycrystalline samples of the spin tube material
[(CuCl$_2$tachH)$_3$Cl]Cl$_2$.  The solid curve in \figref{F-C}
depicts the specific heat of a spin-3/2 periodic AHC composed of
100 spin sites. The specific heat is evaluated by means of a QMC
method employing the ALPS code \cite{ALPS:JMMM07}. As an
effective exchange parameter we used $J_{\text{eff}}=2
J_2/3$. As can be seen in \figref{F-C}, the QMC result
reproduces the experimental data very well in the region $T\leq
0.5$.  This means that even for $J_2^{'}=0$, when the parameter
$(J_2^{'}-J_2)$ is definitely not small, the low-energy physics
of Hamiltonian (\ref{E-2-1}) is described by the spin-3/2
AHC. Therefore, it turns out that the theoretically predicted
biquadratic exchange term \cite{LNM:PRB04,FLP:PRB06} plays no
role in the experimentally interesting region of the phase
diagram characterized by the dimensionless parameter
$J_1/J_2=0.46$.

%===================    figure   =================================
%===================    figure   =================================

Turning to the extremely low-temperature regime, we expect the
universal specific-heat behavior of the Tomonaga-Luttinger
liquid
%--------------------------------------------------------
\begin{eqnarray}
\label{E-4-1}
\frac{C(T)}{Nk_B}=\frac{\pi c T}{9v_s}
\ ,
\end{eqnarray}
%--------------------------------------------------------
where the constant $c$ stands for the so-called topological
charge ($c=1$ for a Tomonaga-Luttinger liquid state) and $v_s$
is the velocity of the gapless spin excitations.  As clearly
seen in \figref{F-C}, already for $T\leq 0.5$~K the measured
specific heat coincides with the QMC results and nicely
extrapolates towards the universal behavior repesented by
Eq.~(\ref{E-4-1}). The latter observations strongly imply that
the spin tube compound [(CuCl$_2$tachH)$_3$Cl]Cl$_2$ is
characterized by a gapless Tomonaga-Luttinger liquid ground
state. In a recent report, Nuclear Magnetic Resonance
measurements also indicate a gapless spin state in the same
material based on estimates for the extremely low-temperature
part of the magnetic susceptibility \cite{FSK:JPCS09}.

To explain the experimental data at intermediate temperatures around
2~K, we use the established structure of the low-lying excitation
spectrum, compare \figref{F-B}. As a rough approximation, one
may take $E_{1,2}(k_x)\approx \Delta$ and use the well-known
expression for the specific heat of a two-level system (with the
assumption that the excited level has twice the weight of the
ground state, i.e., $r=2$ in the following expression).  Since
the exact density of states is not known in the thermodynamic
limit, we use an overall parameter $A$ in order to fix the
height of the Schottky peak:
%--------------------------------------------------------
\begin{eqnarray}
\label{E-4-6}
\frac{C}{Nk_B}=A\frac{r(\Delta/T)^2\exp (\Delta/T)}{[\exp
(\Delta/T)+r]^2}
\ .
\end{eqnarray}
%--------------------------------------------------------
Notice that the position of the peak does not depend on the
value of $A$.  The expression for $C(T)$, Eq.~(\ref{E-4-6}),
with $\Delta/k_B=5.32$ K is plotted in \figref{F-D} by a thick
curve together with the experimental data. One observes that it
reproduces very well not only the position of the peak
($T_m\approx 2$~K) but also the behavior of the specific heat
for $T<T_m$ down to $T \approx 0.7$ K, where the contribution
from the gapless branch $E_0(k_x)$ becomes important. In
addition, in \figref{F-D} we also show the specific heat that
results from complete diagonalizations of a few finite-size
periodic clusters. As easily seen, the overall agreement is
good, in spite of the pronounced finite-size effects.
%===================    figure   =================================
%===================    figure   =================================

Turning to higher temperatures, a few words are necessary to
expain the discrepancies between the experimental and
numerical results for $C(T)$. Notice that already at $T\approx
2$~K the phonon contribution begins to dominate the specific
heat.  Since the detailed phonon spectral density is unknown,
the raw experimental data for $C(T)$ was -- as usually --
corrected by subtracting a reasonable Debye-like specific heat
contribution \cite{YNO:NP08}.

\emph{Conclusion}--We demonstrated that the low-temperature
specific heat behavior of the spin tube compound
[(CuCl$_2$tachH)$_3$Cl]Cl$_2$ suggests a Tomonaga-Luttinger
liquid type ground state for this material, corresponding to an
effective spin-3/2 antiferromagnetic Heisenberg chain
characterized by the short-ranged exchange-coupling constant
$J_{\text{eff}}=2J_2/3$.  On the other hand, we argued that the
main contribution to the observed Schottky-type peak around
$T\approx 2$ K comes from the lowest-lying gapped magnon-type
excitations resulting from the internal degrees of freedom of
the composite rung spins.

%%%%%%%%%%%%%%%%%%%%%%%%%%%%%%%%%%%%%%%%%%%%%%%%%%%%%%%%%%%%%%%%%%%%%%%%
\emph{Acknowledgments}--Computing time at the Leibniz Computing
Center in Garching is gratefully acknowledged. We also thank
Andreas Honecker and David Johnston for fruitful discussions as
well as Tao Xiang for explaining his transfer matrix results to
us.  J.~S. is grateful to Andreas L{\"a}uchli for advising how
to run the ALPS code \cite{ALPS:JMMM07}. This work was supported
by the DFG (FOR~945, SCHN~615/13-1) and the Bulgarian Science
Foundation under the Grant No. DO02-264. H.~N. is supported by
Kakenhi No. 20244052 from JSPS. Part of the ED results was
obtained with spinpack.

%\bibliography{/home/schnack/tex/bibtex/js-own,/home/schnack/tex/bibtex/js-mag,/home/schnack/tex/bibtex/js-mis}
%\bibliography{/Users/jschnack/tex/bibtex/js-own,/Users/jschnack/tex/bibtex/js-mag,/Users/jschnack/tex/bibtex/js-mis}

\end{document}